\documentclass{article}
\usepackage{graphicx,psfig,epsfig}
\def\title{\begin{center}\Large\bf}
\def\author(s){\vspace{0.3cm}\large\rm}
\def\text{\end{center}}

\begin{document}

\title
Seeing Measurements at Skinakas Observatory using the DIMM method

\author(s)
P. Boumis$^{\rm 1,2}$, A. Steiakaki$^{\rm 2}$, F. Mavromatakis$^{\rm 1,2}$, 
G. Paterakis$^{\rm 1}$, and  \\ J. Papamastorakis$^{\rm 1,2}$

$^{\rm 1}${\it University of Crete, P.O Box 2208, GR-71003, Heraklion, Greece}\\
$^{\rm 2}${\it Foundation for Research and Technology-Hellas, P.O. Box 1527,\\
GR-71110 Heraklion, Greece}\\
\text

\vspace{0.3cm}

\large

\section*{Abstract}
We present preliminary results from a study of the seeing at Skinakas
Observatory in Crete, Greece. The measurements have been made during
the years 2000 and 2001 using a two-aperture Differential Image Motion
Monitor (DIMM). The results of both campaigns are extremely promising
with the median seeing value being 0.$''$64 and 0.$''$69,
respectively.

\section{Introduction}
The knowledge of the atmospheric turbulence, which limits the
telescope's angular resolution, is very important for the correct
design and work of Adaptive Optics systems. It is well known that when
light is coming from a star, it crosses several turbulence atmospheric
layers which results to beam degradation at the telescope focus. The
common way to characterize image degradation, is to measure the full
width at half-maximum (FWHM) intensity of a star (in arcsec) at the
focus position, which is actually its angular diameter that we call
$``$seeing$"$ in astronomy. The method which is accepted as the most
accurate for seeing measurements is called $``$the Differential Image
Motion Monitor (DIMM)$"$, which is the study of the differential
motion of the image of a star. The same method with small differences
was used at Skinakas Observatory to study the changes of the
atmospheric turbulence in the area, hence the site's observational
quality. In section 2, there is a description of the DIMM Optical
configurations. Section 3 is dedicated to explain the method, while
the observations and results are given in section 4 and 5,
respectively.

\section{DIMM Optical Configuration}
The configuration used for the DIMM measurements at Skinakas
Observatory is similar with the DIMM used by Wood et al (1995) and
Vernin \& Munoz-Tun\'{o}n (1995). It is presented schematicly in
Fig.~1, while, the technical equipment which have been chosen in the
design and construction of the Skinakas DIMM are as follows: (1) Meade
LX200 12" Telescope, (2) Equatorial Mount, (3) CCD camera SBIG ST-4,
(4) Mask with two holes, (5) PC computer, (6) DIMM code, (7) Pointing
program and (8) Autoguiding program. Also, the characteristics of the
telescope, the CCD camera as well as the optical configuration (shown
in Fig. 1) of the DIMM are given in Table 1.

\section{DIMM Operation}
The DIMM technique is to measure the relative motion of two images of
the same star and based on this motion, the size of the holes and their
separation, the seeing can be derived. The advantage of the
differential method is that eliminates erratic motion of the
telescope, since it measures the angular differences between the two
images and it is not affected by any other motion. The idea
is that the light from a star passing through the two
apertures, travels through slightly different atmospheric conditions
procuding a tilt in the wavefront of one compared to the other. This
results in a slight variation in the separation of the two images.

\begin{figure}
\centering
\mbox{\epsfclipon\epsfxsize=2.8in\epsfbox[0 0 480 382]{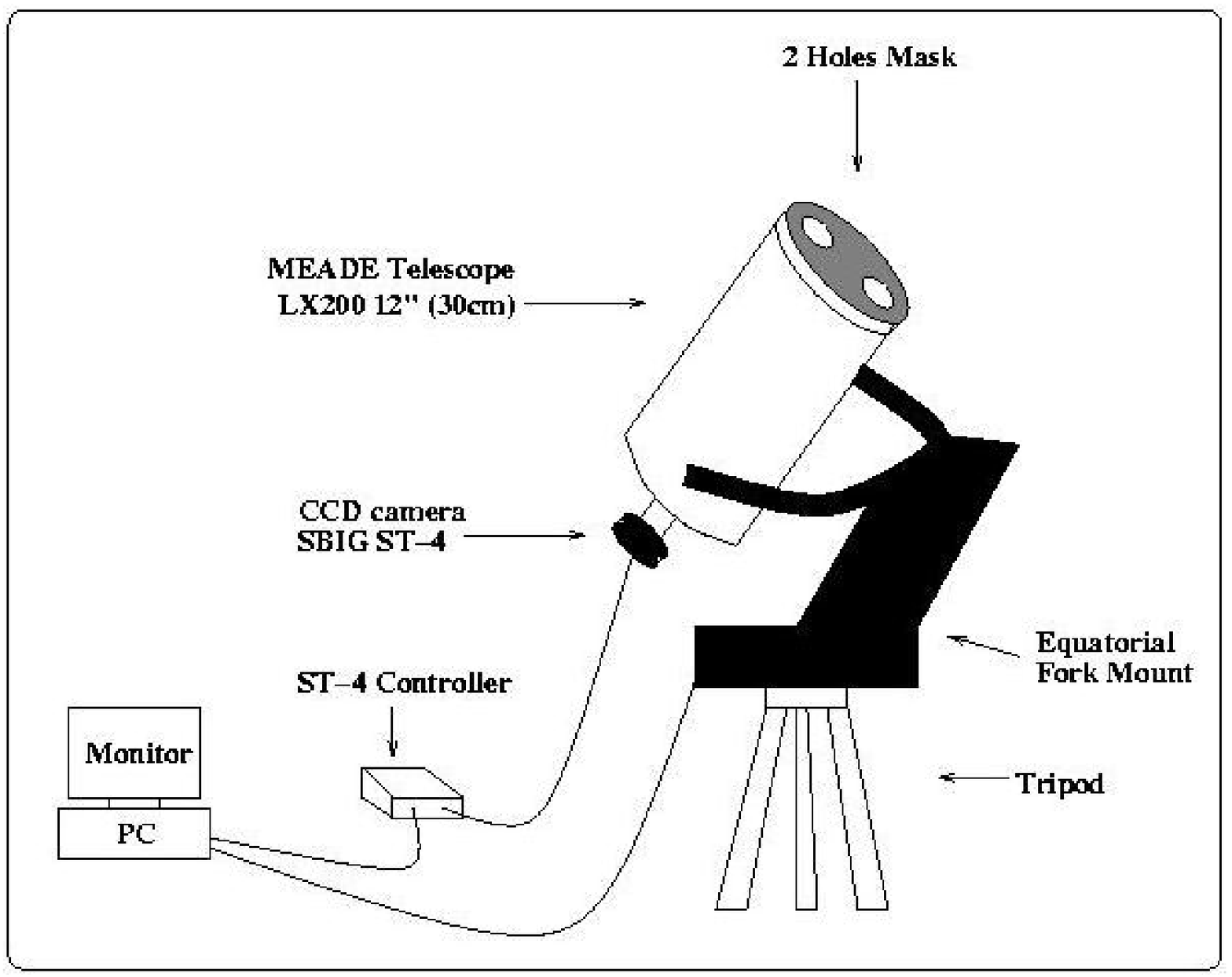}}
\mbox{\epsfclipon\epsfxsize=2.8in\epsfbox[0 0 437 344]{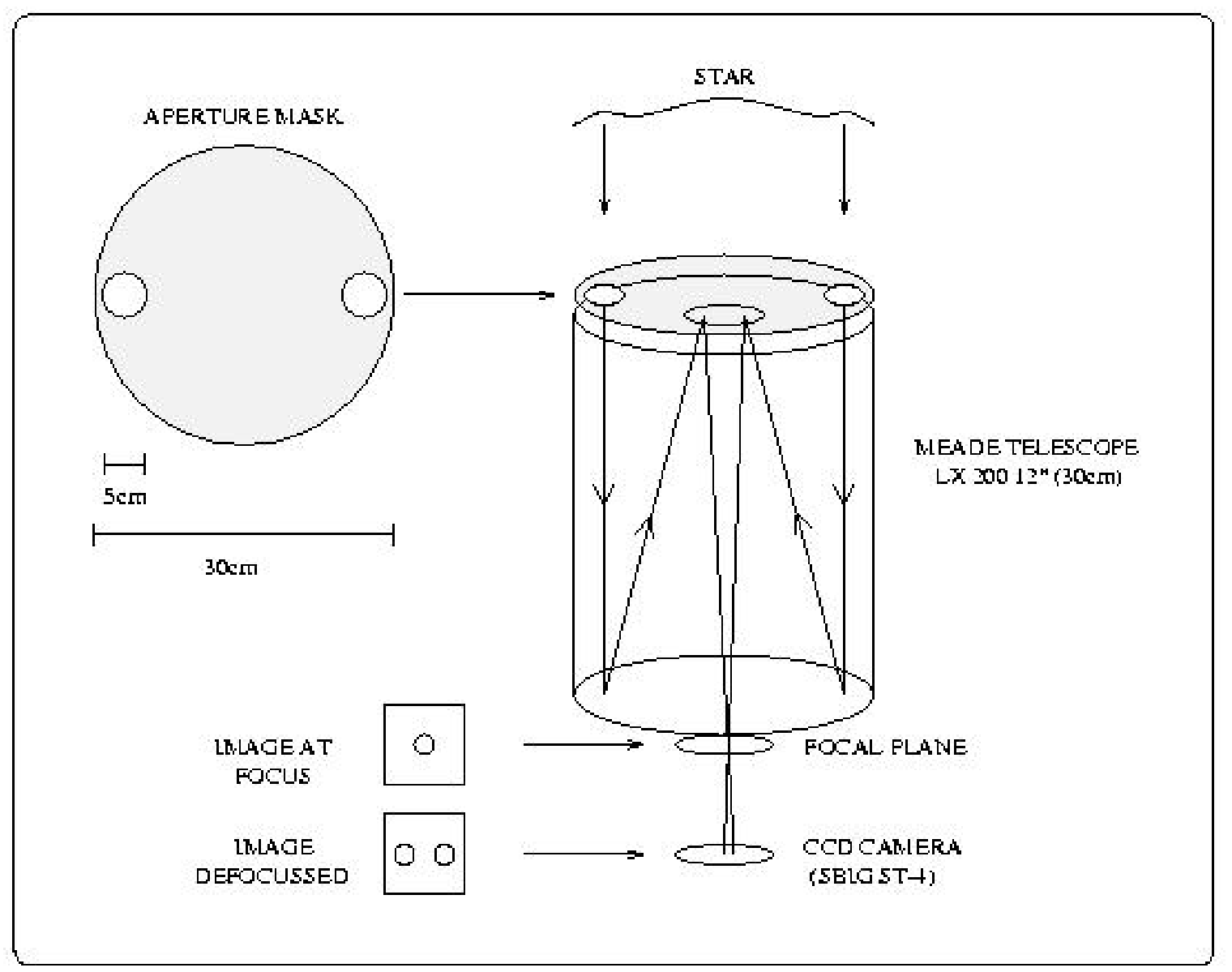}}
\\ {{\bf Figure 1.} A sketch with the Dimm and the Optical Configuration}
\label{fig1}
\end{figure}

\begin{center}
\begin*{}
\centering
\begin{tabular}{c c}
\multicolumn{2}{c}{{\bf Table 1.} DIMM Configuration}\\
\multicolumn{2}{c}{}\\
\hline
\multicolumn{2}{c}{{\bf Meade LX200 12$"$~Specifications}}\\
\hline
Optical design & Schmidt-Cassegrain \\
Clear aperture & 305 mm (12")\\
Focal length & 3048 mm \\
Limiting visual mag & $\sim$15 \\
Image scale & 1.14 arcmin/mm \\
\hline
\multicolumn{2}{c}{{\bf SBIG ST-4 CCD Specifications}}\\
\hline
Pixel array & 192$\times$164 pixels$^{2}$\\
Total pixels & 31,000\\
Pixel size & 13.75$\times$16 microns$^{2}$\\
Exposure time & 0.01 to 300 seconds\\
\hline
\multicolumn{2}{c}{{\bf Optical System Specifications}}\\
\hline
Field of view & 2.98$\times$2.96 arcmin$^{2}$\\
Pixel size & 0.93$\times$1.08 arcsec$^{2}$\\
\hline 
\end{tabular}
\end*{} 
\end{center}

In order to get the two images at the correct focus, a thin optical
wedge need to be used over one of the apertures, to produce a small
deflection in one of the beams. At Skinakas Observatory we used a
different method, to avoid using a wedge. The telescope was set
slightly out of focus, which separates the two images, whithout
causing any problem at the image quality in the appropriate level,
since it is not affect the quality of the individual star images as
the f/ratio is very high. In the case of the MEADE LX200 12$"$, the
telescope's f/ratio is f/10, while that of the two apertures is
f/60. As a result, the image is not significantly blurred by a slight
defocussing of the telescope.

The DIMM code is based on the mathematical equations used by ESO/DIMM
(Sarazin \& Roddier 1990). The parameters which must be specified in
the code are: 

\begin{enumerate}
\item The diameter of the two holes in the mask (5 cm)
\item The separation of the two holes - centre to centre (24.7 cm)
\item The pixel size of the telescope/CCD camera combination in both x, y
dimensions ($\sim$1 arcsec) 
\end{enumerate}

while, the input data to start running the program are: 

\begin{enumerate}
\item Coordinates of the observed star 
\item Exposure time for the seeing measurements
\item Exposure time for testing images 
\item A parameter which is correlated with the magnitude of the star
\end{enumerate}

The basic part of the program is to calculate exactly the centres of
the two images of the star in each exposure. After two subsequent
exposures, any variation between the two images is calculated in both
directions (parallel and perpendicular to aperture alignment),
while after a specific number of exposures the longitudinal $\sigma$(l) and
transverse $\sigma$(t) variance of the differential image motion can be
deduced. The latter, in conjuction with the equations (1) and (2)
estimate the Fried's parameters r$_{o}$(l) and r$_{o}$(t), which characterize
the atmospheric condition at the site of observations. Finally, using
the equations (3) and (4) the "seeing" can be derived.

\begin{equation}
\sigma(\rm l)^{2} = 2\lambda^{2} [0.179 / D_{\rm HOLE}^{1/3} - 0.097 / d_{\rm SEP}^{1/3}
] r_{\rm o}(l)^{-5/3}
\end{equation}                           

\begin{equation}
\sigma(\rm t)^{2} = 2\lambda^{2} [0.179 / D_{\rm HOLE}^{1/3} - 0.145 / d_{\rm SEP}^{1/3}
] r_{\rm o}(t)^{-5/3}
\end{equation}                           
where,
\newline $\sigma$(l), $\sigma$(t) : longitudinal and transverse
(parallel and perpendicular to aperture alignment) variance of
differencial image motion\\
D$_{\rm HOLE}$~: diameter of the holes\\
d$_{\rm SEP}$~: separation of the two holes (centre to centre)\\ 
r$_{\rm o}$~: Fried's parameter\\
$\lambda$~: wavelength (550 nm)

and      
\begin{equation}
{\rm FWHM(l)} = 0.98\lambda / {\rm r_{o}(l)}
\end{equation} 
    \begin{equation}
{\rm FWHM(t)} = 0.98\lambda / {\rm r_{o}(t)}
\end{equation}

\section{Observations \& Results}
The Skinakas DIMM started at the end of April 2000 at the Physics
Department of the University of Crete in Heraklion where all the
apropriate tests took place. A month later it was installed to
Skinakas Observatory (1750 m altitude) and the seeing measurements
started. The latter, took place from the beginning to the end of each
astronomical night in specific dates (dependent to Skinakas staff
availability) from June to September 2000 and from May 2001 up to
now. The place, where the telescope has been installed, is according
to our requirements (not any thermal source close to the telescope,
the inside-outside temperature difference was not larger than 1$^{\rm
o}$C and continiously normal change of the air inside the dome) so the
seeing that we measured was exaclty that caused by the atmospheric
layers.

Part of the results (see Boumis et al. 2001) of the Skinakas DIMM
measurements are presented in Fig. 2. The seeing is given in
arcseconds in both directions (l,t - y axis) versus the local time
(x-axis). Furthermore, a histogram and the median seeing value for
each night is also given. It must be noted that all measurements are
corrected to zenith. The diagrams show that the seeing usually does
not change rapidly during the observing night (not more than 0.3$''$)
at Skinakas. Of course, there are a few nights where a big difference
occurred (up to 1.5$''$) and the reason was the rapid changes in the
atmospheric conditions. Extremely good seeing values have been
measured (0.4$''$) often, with the best value at
$\sim$0.27$''$. Finally, in Fig. 3, histograms with the seeing
measurements for the year 2000 and 2001 are given, where a median
seeing value of 0.64$''$ and 0.69$''$ have been calculated,
respectively.

\section*{Acknowledgments}
The authors would like to acknowledge support from a "P.EN.E.D." 
program of the General Secretariat of Research and Technology of
Greece. We also thank A. Kougentakis for his assistance during the
observations. Skinakas Observatory is a collaborative project of the
University of Crete, the Foundation for Research and Technology-Hellas
and the Max-Planck-Institut fur Extraterrestrische Physik.

\section*{References}
Boumis P., Mavromatakis F., Steiakaki A., Paterakis G., Papamastorakis
J., 2001, Joint European and National Astronomical Meeting, JENAM
2001, held in Munich, Germany, 10-15 September, 2001, Abstract, p. 219.\\
Sarazin M., Roddier F., 1990, A\&A, 227, 294\\
Vernin J., Munoz-Tun\'{o}n C., 1995, PASP, 107, 265\\
Wood P. R., Rodgers A. W., Russel K. S., 1995, PASA, 12, 97\\

\begin{figure}
\centering
\mbox{\epsfclipon\epsfxsize=1.8in\epsfbox[0 0 518 327]{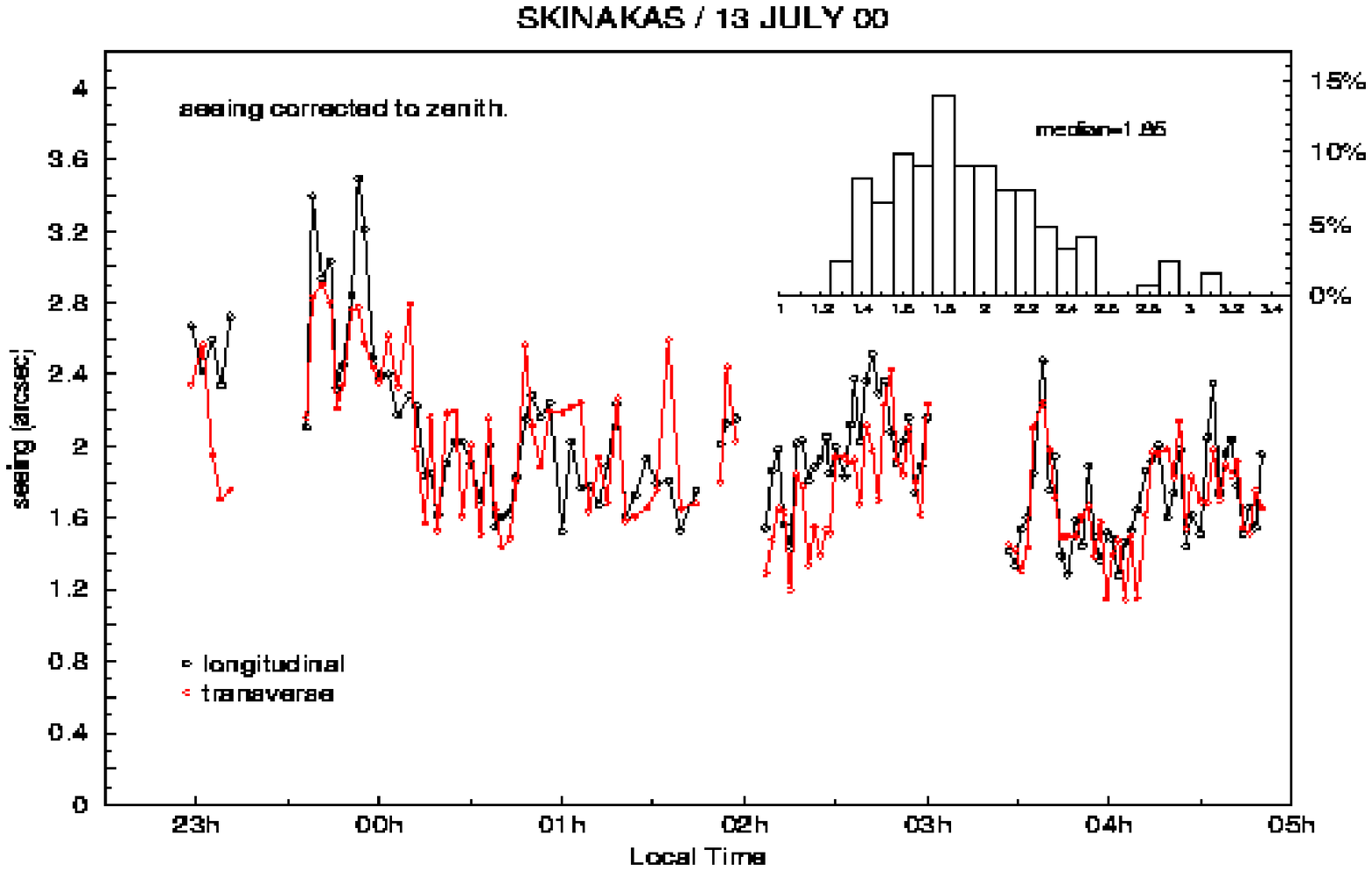}}
\mbox{\epsfclipon\epsfxsize=1.8in\epsfbox[0 0 514 329]{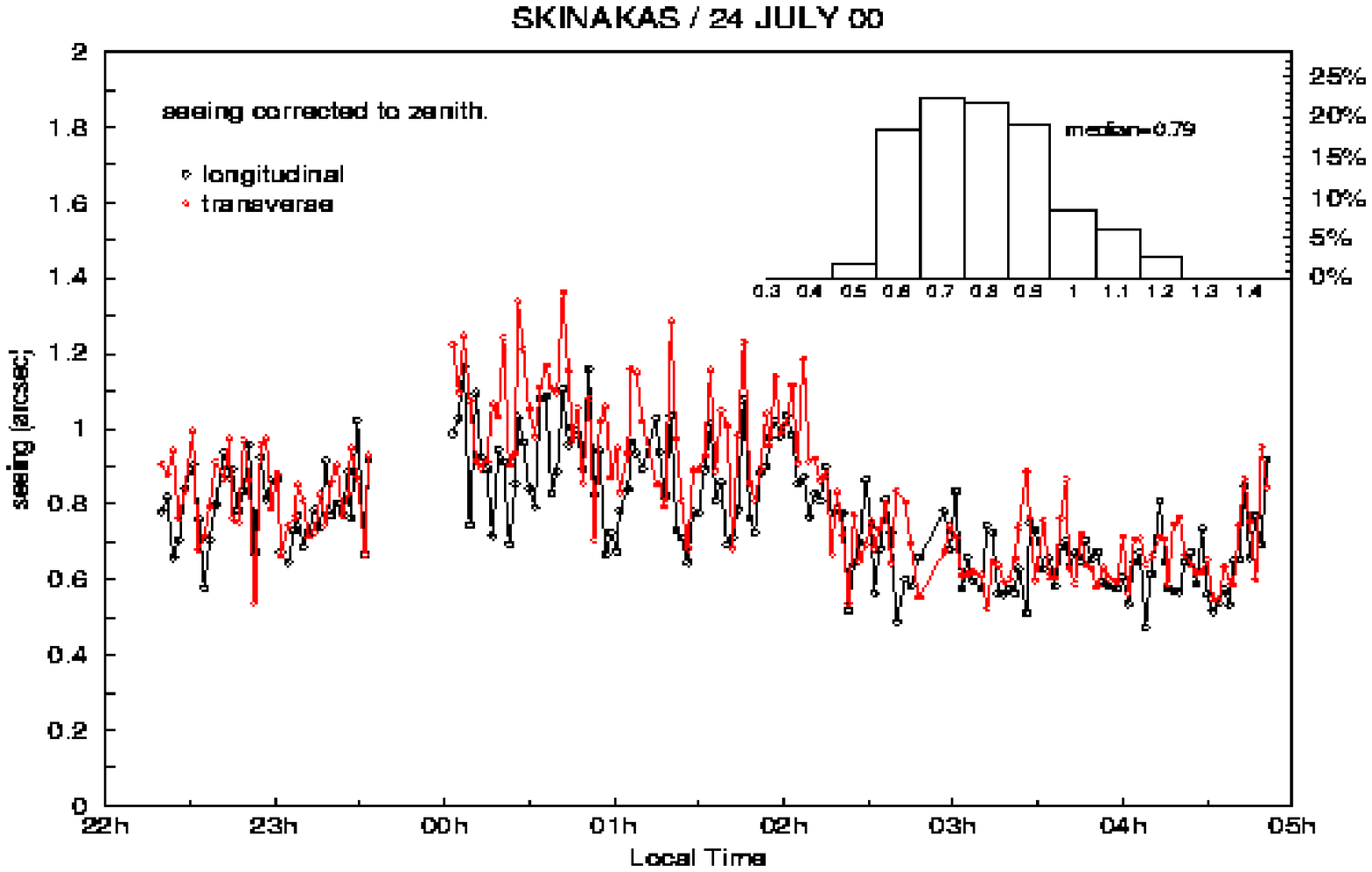}}
\mbox{\epsfclipon\epsfxsize=1.8in\epsfbox[0 0 498 317]{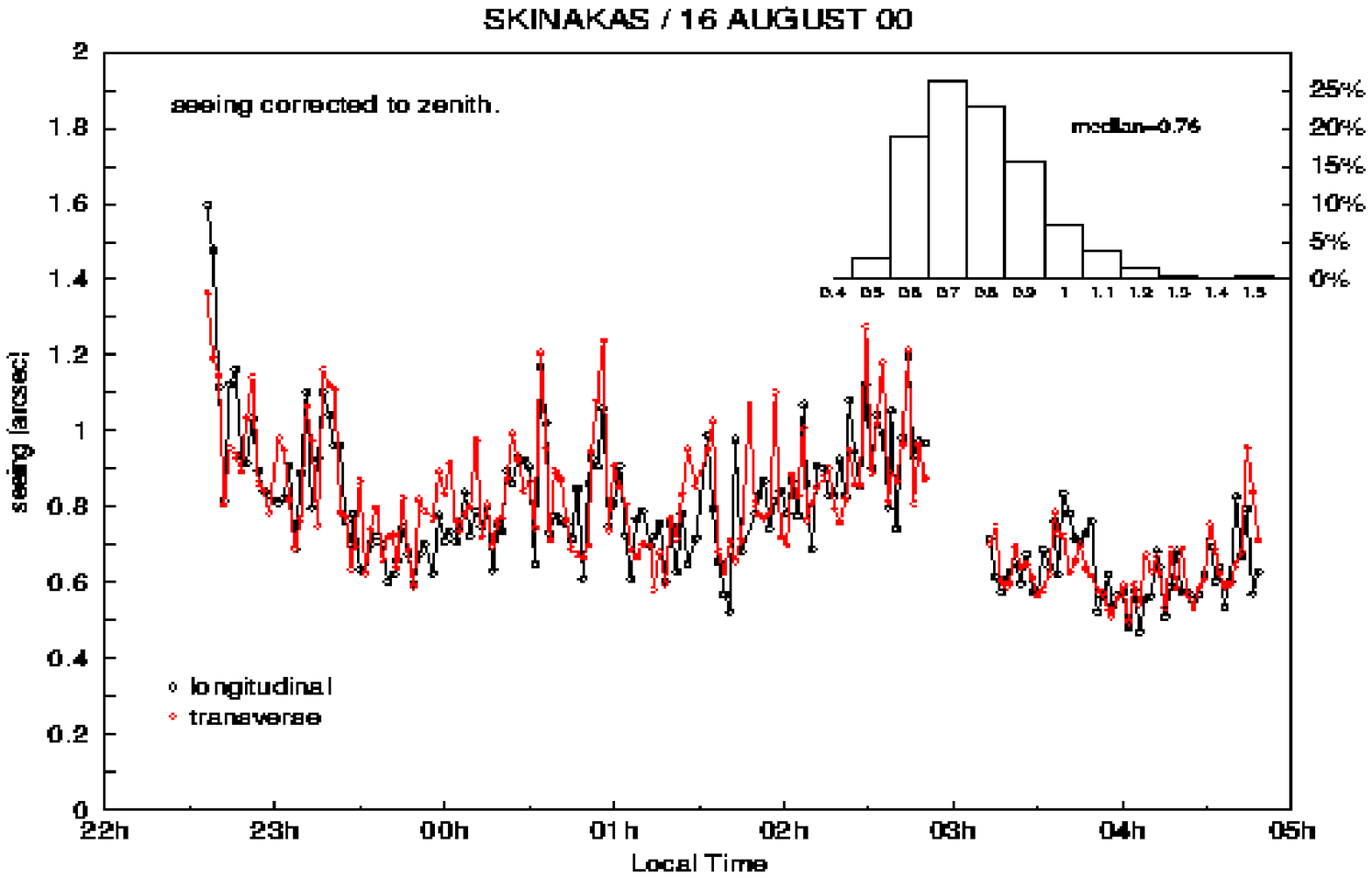}}
\mbox{\epsfclipon\epsfxsize=1.8in\epsfbox[0 0 510 328]{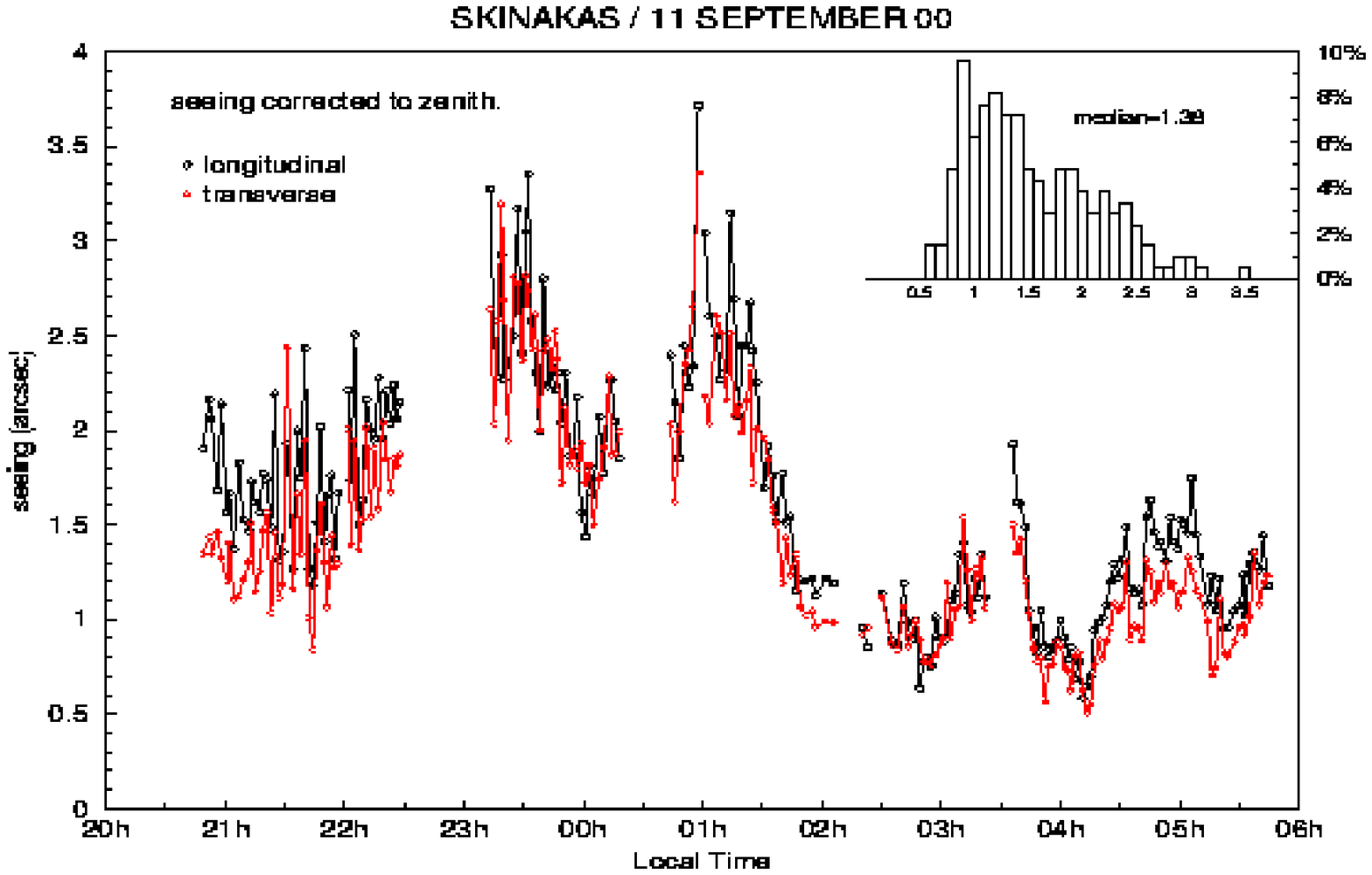}}
\mbox{\epsfclipon\epsfxsize=1.8in\epsfbox[0 0 500 329]{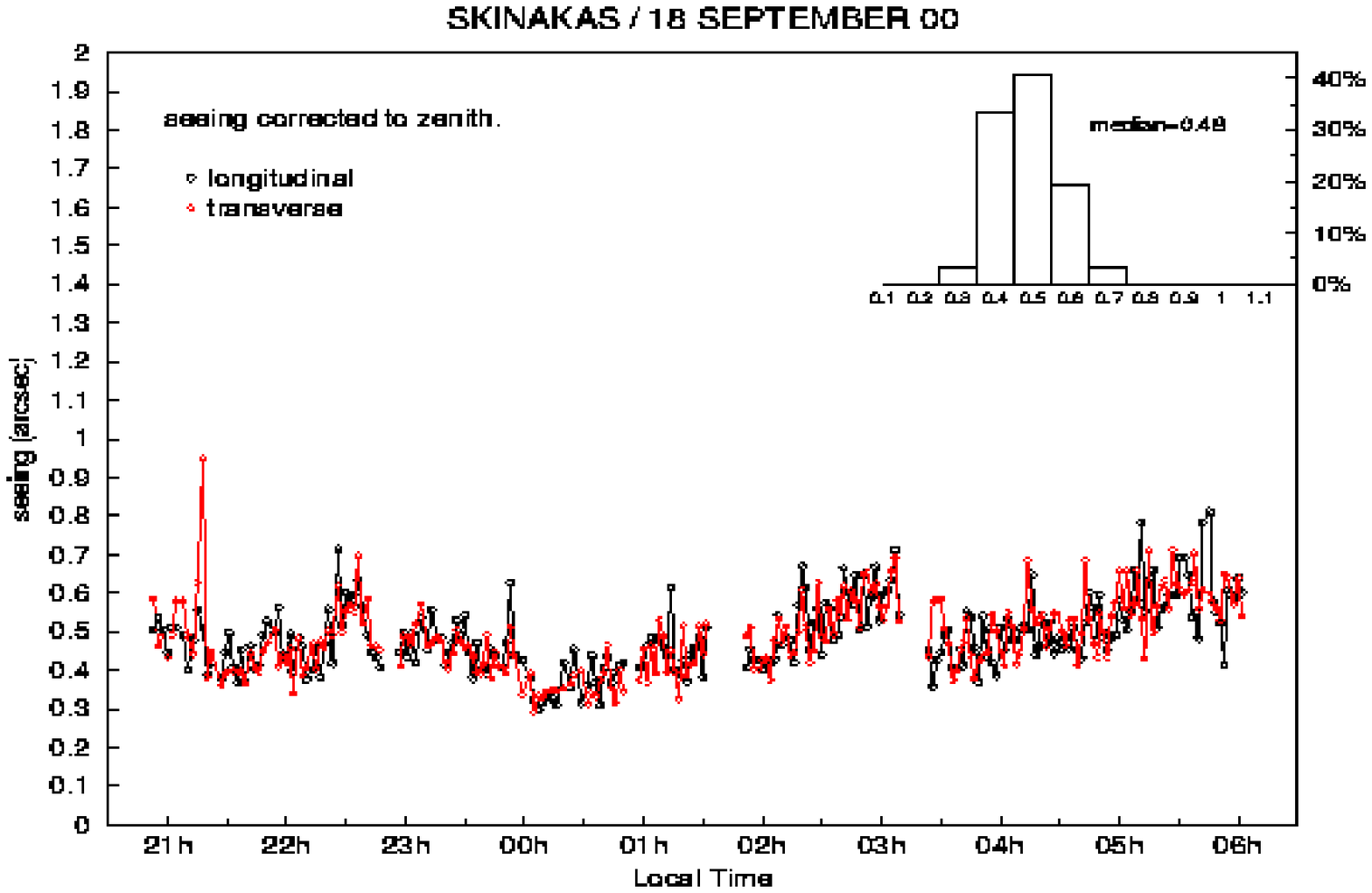}}
\mbox{\epsfclipon\epsfxsize=1.8in\epsfbox[0 0 504 329]{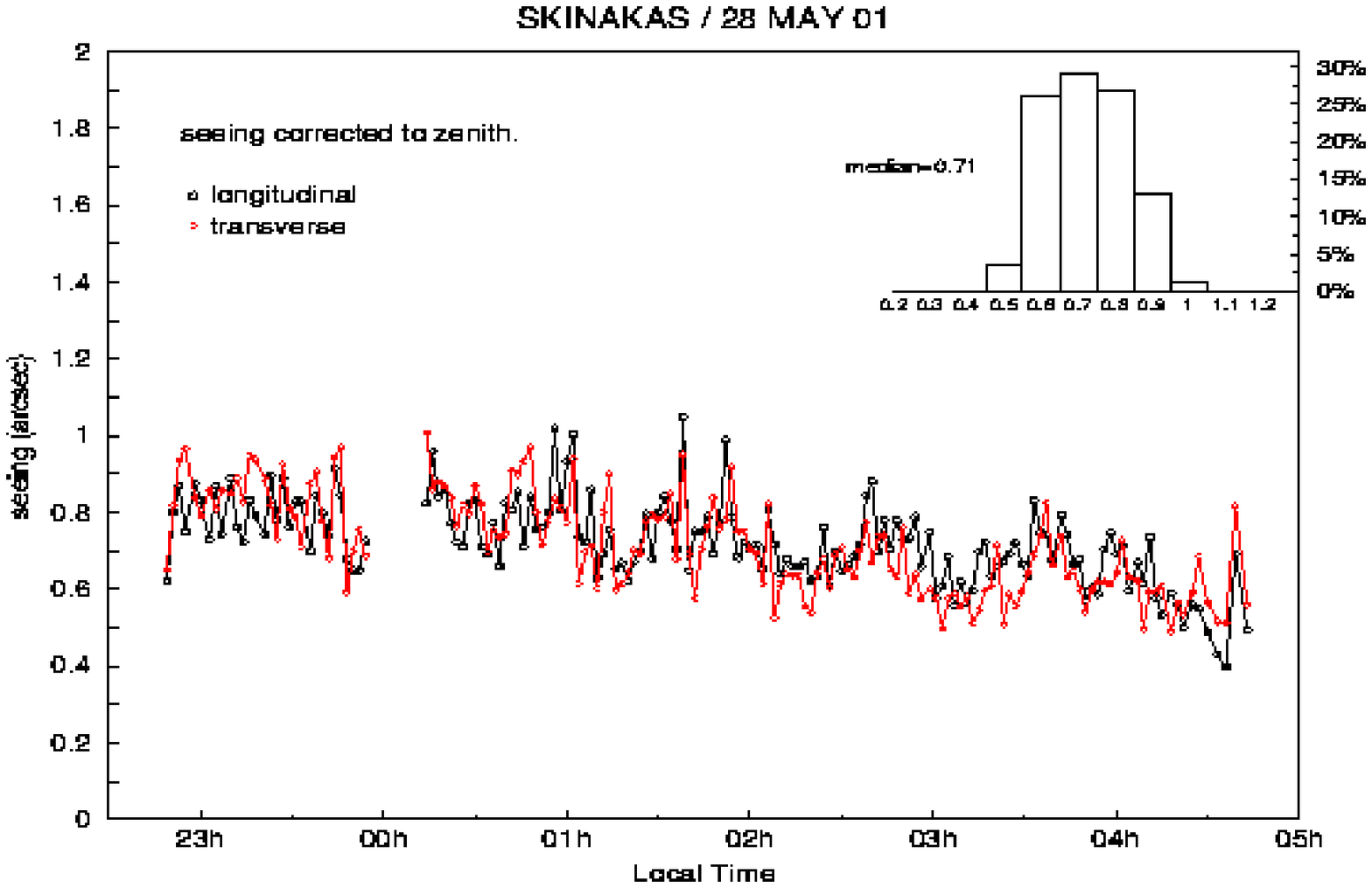}}
\mbox{\epsfclipon\epsfxsize=1.8in\epsfbox[0 0 509 331]{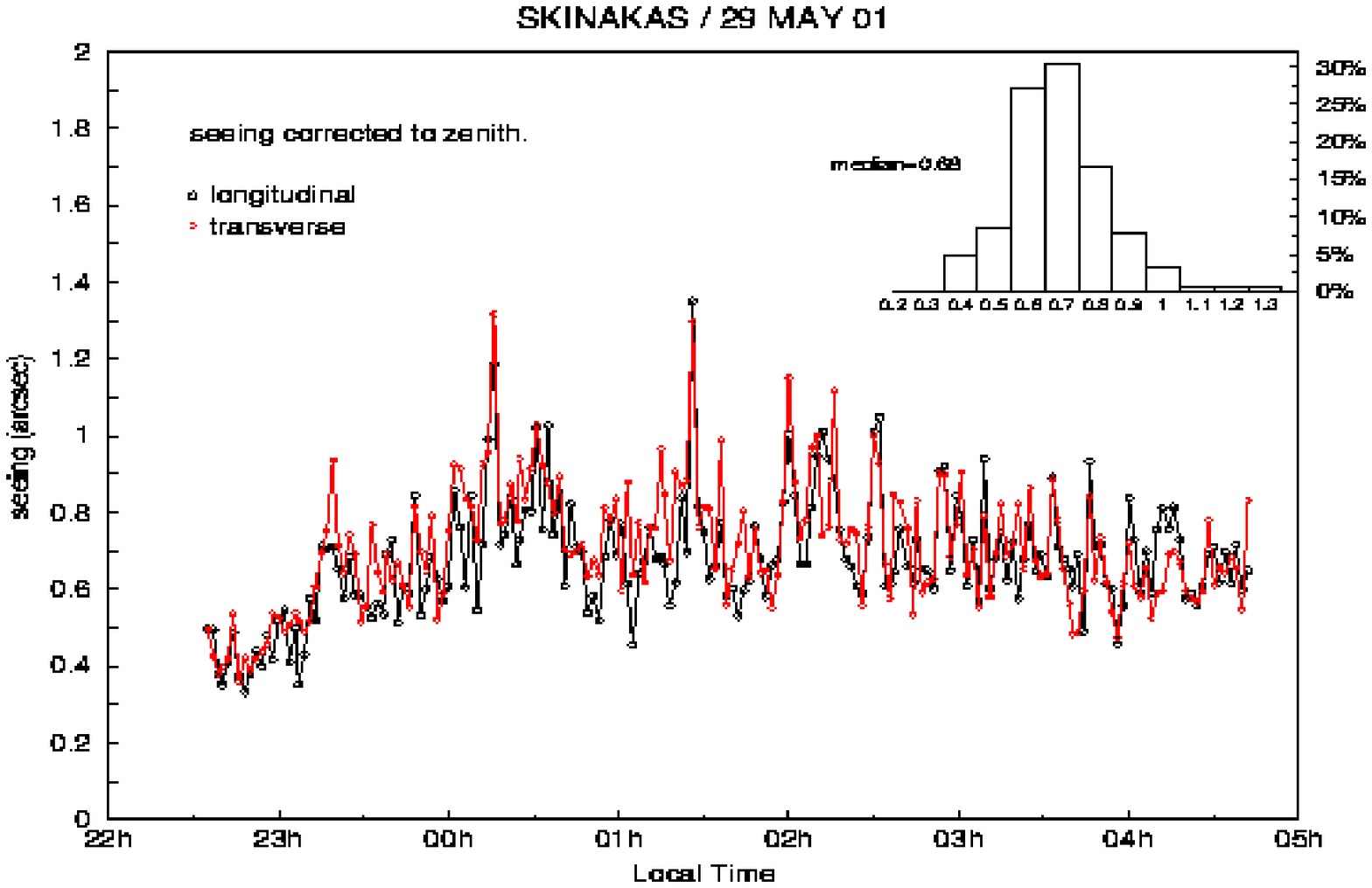}}
\mbox{\epsfclipon\epsfxsize=1.8in\epsfbox[0 0 510 330]{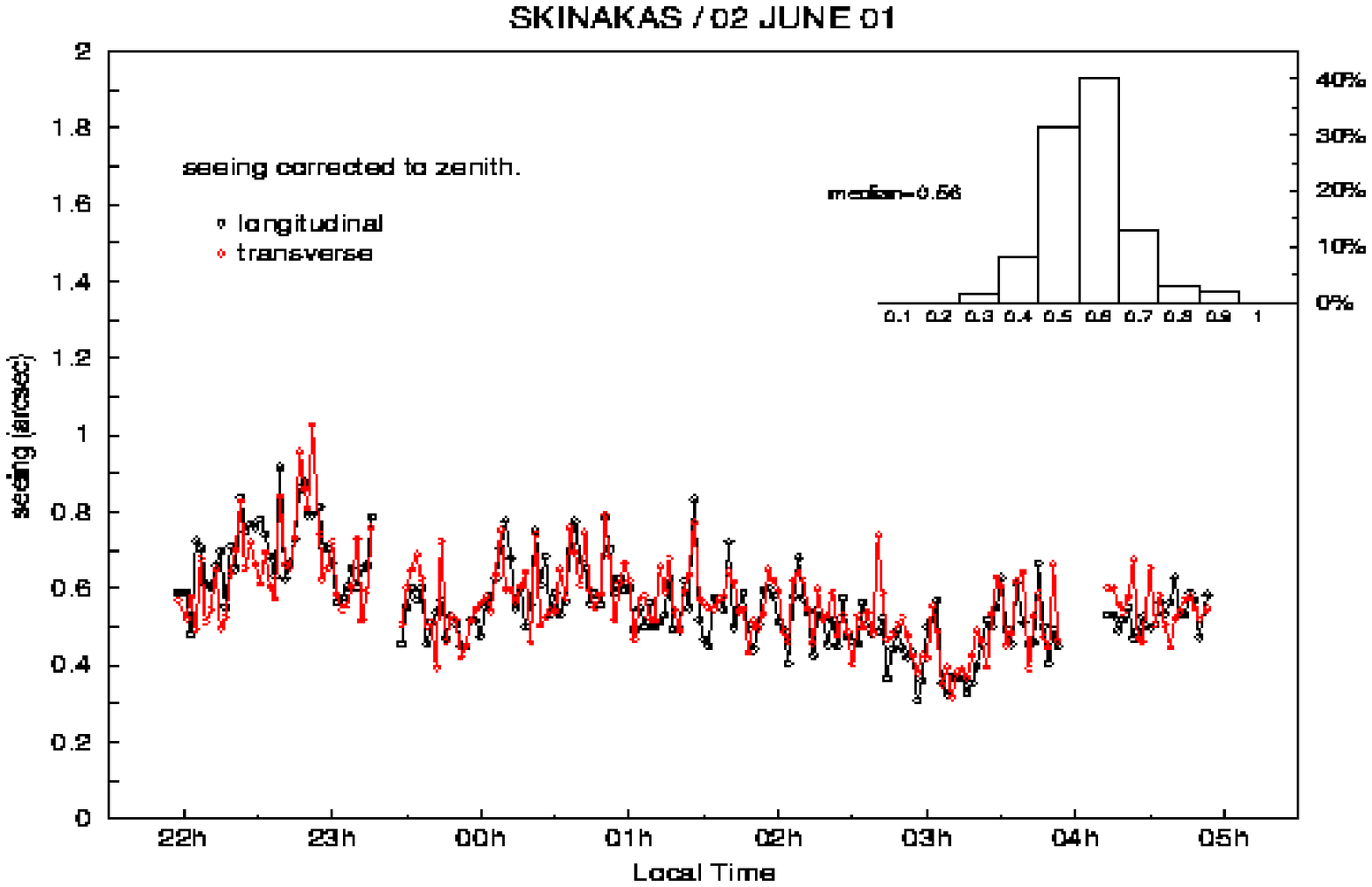}}
\mbox{\epsfclipon\epsfxsize=1.8in\epsfbox[0 0 495 329]{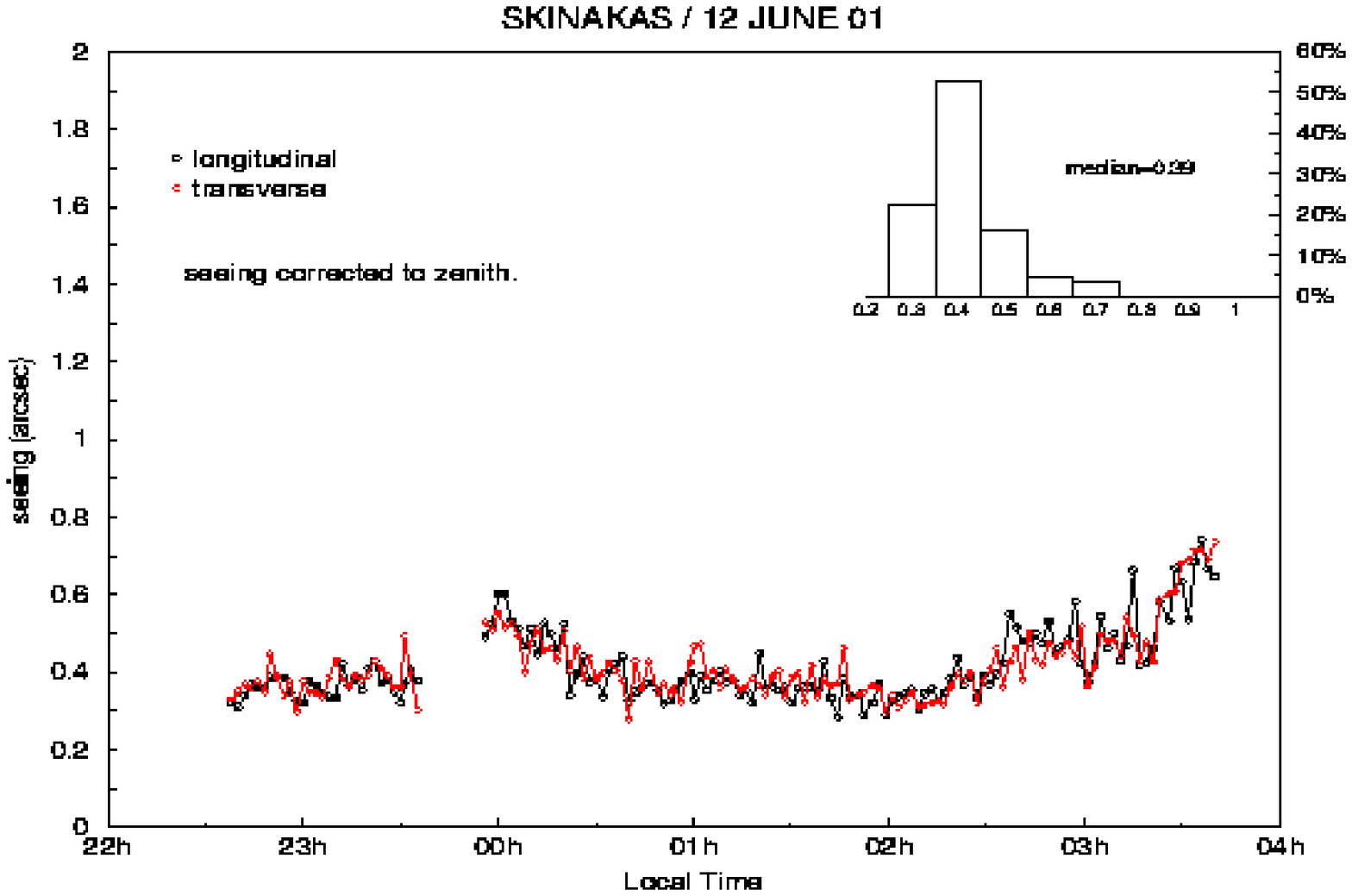}}
\\ {{\bf Figure 2.} Part of the results for the seeing measurements at
Skinakas Observatory for the years 2000 and 2001}
\label{fig2}
\end{figure}

\begin{figure} 
\centering
\mbox{\epsfclipon\epsfxsize=2.7in\epsfbox[0 0 521 372]{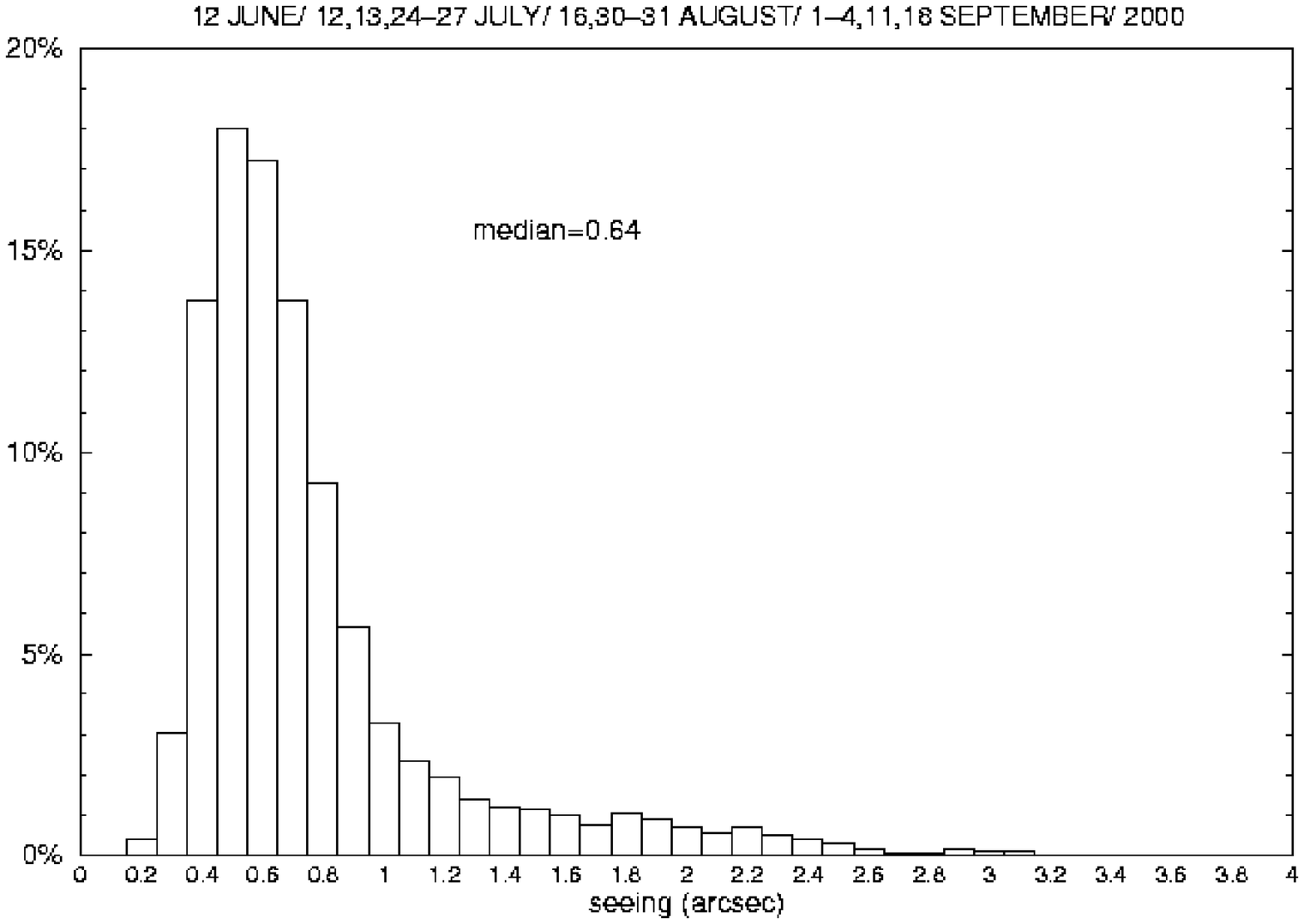}}
\mbox{\epsfclipon\epsfxsize=2.7in\epsfbox[0 0 524 373]{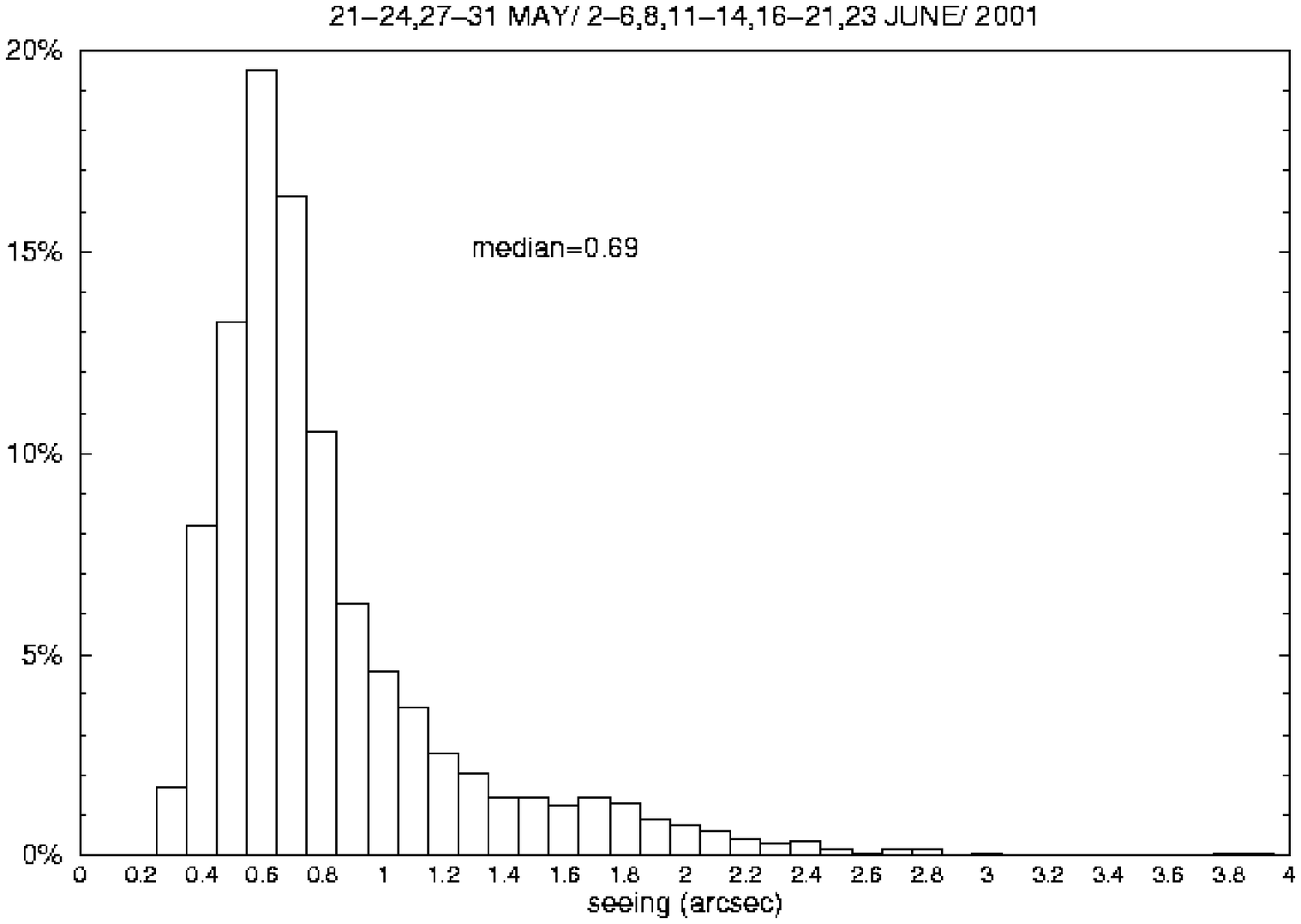}}
\\ {{\bf Figure 3.} Histograms of the seeing distribution at Skinakas
Observatory for the years 2000 and 2001.}
\label{fig3}
\end{figure}

\end{document}